\documentclass[a4paper]{aa}

\usepackage{graphicx,natbib,amsmath}
\usepackage{txfonts}
\usepackage{color}
\usepackage[usenames,dvipsnames]{xcolor}

\newcommand{\bz}{$\langle B_\mathrm{z} \rangle$}
\newcommand{\bm}{$\langle B \rangle$}
\newcommand{\hd}{HD\,75049}
\newcommand{\teff}{$T_{\rm eff}$}
\newcommand{\lgg}{$\log g$}
\newcommand{\vs}{$v_{\rm e}\sin i$}
\newcommand{\kms}{km\,s$^{-1}$}

\newcommand{\figps}[1]{\resizebox{\hsize}{!}{\rotatebox{0}{\includegraphics{#1}}}}

\newcommand{\firrps}[2]{\resizebox{#1}{!}{\rotatebox{270}{\includegraphics{#2}}}}
\newcommand{\figgps}[2]{\resizebox{#1}{!}{\rotatebox{0}{\includegraphics{#2}}}}

\newcommand{\beq}{\begin{equation}}
\newcommand{\eeq}{\end{equation}}

\begin{document}

\title{Magnetic field topology and chemical spot distributions \\ in the extreme Ap star HD\,75049%
\thanks{Based on observations collected at the European Southern Observatory, Chile (ESO programs 084.D-0338, 085.D-0296, 086.D-0240, 088.D-0066, 090.D-0256, 
078.D-0192, 080.D-0170).}}

\author{O.~Kochukhov\inst{1}
  \and N.~Rusomarov\inst{1}
   \and J.~A.~Valenti\inst{2}
   \and H.~C.~Stempels\inst{1}
   \and F.~Snik\inst{3}
   \and M.~Rodenhuis\inst{3}
   \and N.~Piskunov\inst{1}
   \and V.~Makaganiuk\inst{1}
   \and C.~U.~Keller\inst{3}
   \and C.~M.~Johns-Krull\inst{4}
   }

\institute{
Department of Physics and Astronomy, Uppsala University, Box 516, 75120 Uppsala, Sweden
\and
Space Telescope Science Institute, 3700 San Martin Dr, Baltimore MD 21211, USA
\and
Sterrewacht Leiden, Universiteit Leiden, Niels Bohrweg 2, 2333 CA Leiden, The Netherlands
\and
Department of Physics and Astronomy, Rice University, 6100 Main Street, Houston, TX 77005, USA
}

\date{Received 29 September 2014 / Accepted 20 November 2014}

\titlerunning{Magnetic field and chemical spots in the extreme Ap star HD\,75049}
\authorrunning{O.~Kochukhov et al.}

\abstract
{
Intermediate-mass, magnetic chemically peculiar (Ap) stars provide a unique opportunity to study the topology of stellar magnetic fields in detail and to investigate magnetically driven processes of spot formation.
}
{
Here we aim to derive the surface magnetic field geometry and chemical abundance distributions for the extraordinary Ap star \hd. This object hosts a surface field of $\sim$\,30~kG, one of the strongest known for any non-degenerate star.
}
{
We used time-series of high-resolution HARPS intensity and circular polarisation observations. These data were interpreted with the help of magnetic Doppler imaging and model atmospheres incorporating effects of a non-solar chemical composition and a strong magnetic field.
}
{
Based on high-precision measurements of the mean magnetic field modulus, we refined the rotational period of \hd\ to $P_{\rm rot}=4.048267\pm0.000036$~d. We also derived basic stellar parameters, \teff\,=\,$10250\pm250$~K and \lgg\,=\,$4.3\pm0.1$. Magnetic Doppler imaging revealed that the field topology of \hd\ is poloidal and dominated by a dipolar contribution with a peak surface field strength of 39~kG. At the same time, deviations from the classical axisymmetric oblique dipolar configuration are significant. Chemical surface maps of Si, Cr, Fe, and Nd show abundance contrasts of 0.5--1.4~dex, which is low compared with many other Ap stars. Of the  chemical elements, Nd is found to be enhanced close to the magnetic pole, whereas Si and Cr are concentrated predominantly at the magnetic equator. The iron distribution shows low-contrast features both at the magnetic equator and the pole.
}
{
The morphology of the magnetic field and the properties of chemical spots in \hd\ are qualitatively similar to those of Ap stars with weaker fields. Consequently, whatever mechanism forms and sustains global magnetic fields in intermediate-mass main-sequence stars, it operates in the same way over the entire observed range of magnetic field strengths.
}


\maketitle

\section{Introduction}
\label{intro}

Magnetic Ap stars are unique natural laboratories that offer the possibility to study the interaction between a strong magnetic field and astrophysical plasma with unprecedented detail. Ap
stars comprise about 10\% of all intermediate-mass main-sequence stars and possess globally organised magnetic fields with a typical strength of a few kG. These magnetic fields are probably stable fossil remnants from an earlier epoch of stellar evolution \citep{braithwaite:2006}. They play a key role in the hydrodynamics of the outer stellar envelope and govern the structure formation at the stellar surface. The magnetic field introduces an anisotropic component in the radiative diffusion process, which leads to the formation of spots of enhanced or depleted element concentration and an inhomogeneous vertical distribution of chemical elements \citep{michaud:1981,alecian:2010}. Observational manifestations of these chemical structures can be studied in detail \citep[e.g.][]{kochukhov:2004e,kochukhov:2006b}, although comprehensive quantitative theoretical models explaining their formation are still lacking.

It is widely believed that the magnetic field topologies of Ap stars are predominantly axisymmetric and roughly dipolar, with the magnetic axis frequently misaligned relative to the stellar rotational axis. These conclusions are mostly drawn from analyses of the rotational modulation of the so-called integral magnetic observables, such as the mean longitudinal magnetic field and the mean field modulus \citep{landstreet:2000,bagnulo:2002}. At the same time, studies of Ap star magnetic fields based on direct line profile modelling in several Stokes parameters remain relatively rare. Such investigations tend to reveal a more complicated picture of the surface magnetic field distribution. For example, the full Stokes vector studies of bright Ap stars $\alpha^2$~CVn and 53~Cam discovered small-scale horizontal contributions on top of large-scale dipole-like fields \citep{kochukhov:2004d,kochukhov:2010,silvester:2014}. On the other hand, the four Stokes parameter analysis of the cool Ap star HD\,24712 \citep{rusomarov:2014} found no significant departures from a purely dipolar topology.

It is understood that inferring the topology of the Ap star magnetic fields from the circular polarisation profiles alone is bound to miss certain aspects of the field complexity \citep{kochukhov:2002c,kochukhov:2010}. But even these studies frequently require deviations from the canonical oblique dipolar topology to reproduce observations. In some cases, only a moderate distortion of a dipole is advocated \citep{kochukhov:2014}, while in other situations (so far limited to early-B magnetic stars) a complex non-dipolar field is clearly evident from observations \citep{donati:2006b,kochukhov:2011a}. To summarise, the current paucity of detailed polarisation line profile studies of Ap stars does not allow one to draw any firm conclusions about the systematic variation of the surface field morphology with the field strength or with key stellar parameters, such as mass and age. To this end, examining the magnetic field geometries of Ap stars with unusual characteristics is useful in terms of mapping the relevant parameter space and might provide additional clues to the problem of the field origin and evolution.

Magnetic fields of Ap stars have a well-established lower limit of $\sim$\,300~G for the polar strength of the dipolar component \citep{auriere:2007}. Most Ap stars have fields in the 1--3~kG range. At the same time, the field strength distribution has a long tail extending to several tens of kG. The late-B star HD\,215441 with the 34~kG mean field modulus \citep{babcock:1960} remains the record-holder among magnetic A and B stars. Another B-type star, HD\,137509, exhibits Zeeman splitting consistent with the field modulus of 29~kG \citep{kochukhov:2006c}. Finally, a much cooler Ap star \hd\ with a field modulus variation of
between 24.5 and 29.5~kG over its $\approx$\,4~d rotation period was discovered by \citet{freyhammer:2008} and subsequently analysed by \citet{elkin:2010}. Apart from the modelling of Zeeman-resolved Stokes $I$ metal line profiles in HD\,215441 by \citet{landstreet:1989}, none of these extremely magnetic objects were studied with high-resolution time-series spectroscopy or polarimetry. Consequently, the magnetic field geometries of all three stars remain essentially unknown beyond the schematic low-order multipolar models derived from the phase curves of integral magnetic observables.

In this paper we investigate in detail the magnetic field topology of \hd\ based on the rotational phase variation of very high-resolution intensity and circular polarisation profiles of spectral lines. This is the first analysis of this kind for such an extreme main-sequence magnetic star. In addition to characterising the stellar magnetic field, we present results of mapping the surface distribution of chemical elements, refine the rotational period, and derive fundamental stellar parameters using model atmospheres that fully account for the chemical peculiarities and the strong magnetic field.

\section{Observational data}
\label{obs}

\begin{table}[!th]
\centering
\caption{Journal of HARPS observations of \hd. 
\label{tbl:obs}}
\begin{tabular}{cccccr}
\hline\hline
UT date & HJD & Phase & S/N & Stokes \\
\hline
2010-01-08 & 2455204.8214 & 0.7454 & 136 & $IV$ \\
2010-01-10 & 2455206.7875 & 0.2311 & 144 & $IV$ \\
2010-01-11 & 2455207.7753 & 0.4751 & 146 & $IV$ \\
2010-01-13 & 2455209.8043 & 0.9763 & 146 & $IV$ \\
2010-01-14 & 2455210.7909 & 0.2200 & 131 & $IV$ \\
2010-04-29 & 2455316.4739 & 0.3258 &  79 & $IV$ \\
2010-04-30 & 2455317.4701 & 0.5718 & 111 & $IV$ \\
2010-05-03 & 2455319.5644 & 0.0892 & 120 & $IV$ \\
2011-02-09 & 2455601.7825 & 0.8025 & 141 & $IV$ \\
2011-02-10 & 2455602.7695 & 0.0463 & 143 & $IV$ \\
2011-02-17 & 2455609.6785 & 0.7530 & 221 & $I$ \\
2011-02-18 & 2455610.6482 & 0.9925 & 207 & $I$ \\
2012-03-28 & 2456014.5916 & 0.7743 & 127 & $IV$ \\
2012-03-29 & 2456015.5662 & 0.0150 & 171 & $IV$ \\
2012-03-30 & 2456016.5819 & 0.2659 & 107 & $IV$ \\
2012-03-31 & 2456017.5481 & 0.5046 & 121 & $IV$ \\
2013-02-22 & 2456345.5633 & 0.5307 & 106 & $IQ$ \\
2013-02-22 & 2456345.6263 & 0.5462 &  98 & $IU$ \\
2013-02-24 & 2456347.5494 & 0.0213 & 139 & $IQ$ \\
2013-02-24 & 2456347.6204 & 0.0388 & 151 & $IU$ \\
2013-02-25 & 2456348.5530 & 0.2692 & 152 & $IQ$ \\
2013-02-25 & 2456348.6241 & 0.2868 & 159 & $IU$ \\
2013-02-27 & 2456350.5459 & 0.7615 & 149 & $IQ$ \\
2013-02-27 & 2456350.6169 & 0.7790 & 151 & $IU$ \\
\hline
\end{tabular}
\tablefoot{The columns give the UT date at mid-observation, the corresponding Heliocentric Julian date, rotational phase, the signal-to-noise ratio per pixel measured at $\lambda=5200$~\AA, and the Stokes parameters obtained.}
\end{table}

High-resolution spectropolarimetric observations of \hd\ were collected at the ESO 3.6 m telescope using the HARPS spectrometer \citep{mayor:2003} and the HARPSpol polarimetric unit \citep{piskunov:2011,snik:2011}. A total of 24 observations were acquired during five observing runs carried out over the period of 2010--2013. Spectropolarimetric observations consisted of sequences of sub-exposures (typically $4\times 1000$--1500~s) obtained with different orientations of the quarter- and half-wave retarder plates for the circular and linear polarimeters. Each of these sequences provided an intensity (Stokes $I$) spectrum and one of the other Stokes parameters (either $V$, $Q$ or $U$) with a resolving power of $\lambda/\Delta\lambda\approx110\,000$ and a wavelength coverage 3780--6910~\AA\ with a 80~\AA\ gap at $\lambda=5290$~\AA. In total, 14 Stokes $V$ and 4 Stokes $QU$ observations were obtained, the latter mostly for exploratory purposes. Table~\ref{tbl:obs} provides an observing log of our HARPSpol data set of \hd.

The typical peak signal-to-noise (S/N) ratio of the data is 100--200 per 0.8~\kms\ pixel in the extracted spectrum as measured at $\lambda\approx5200$~\AA. For this S/N the Stokes $QU$ spectra show no linear polarisation signatures suitable for a detailed line profile analysis. Consequently, only the Stokes $I$ spectra from the linear polarisation observations were considered in the analysis described below.

A detailed account of the HARPSpol observing procedures and the key steps of data reduction, such as the optimal extraction of \'echelle spectra and calculation of the Stokes parameters from a sequence of polarimetric sub-exposures, was given by \citet{rusomarov:2013}. In the present paper we use the exact same observing procedures and reduction techniques.

To extend the baseline for determining the stellar rotational period, we downloaded from the ESO Archive the UVES intensity spectra of \hd\ described by \citet{elkin:2010}. These data, comprising 13 spectra collected in 2007--2008, were processed with the {\sc reduce} pipeline \citep{piskunov:2002} and were used here solely to determine the mean field modulus from the \ion{Fe}{ii} 5018.44~\AA\ line. The resolution and wavelength coverage of the UVES spectra of \hd\ are similar to the HARPS data, while the S/N is about 250 per 1.2~\kms\ pixel.

\section{Stellar parameters}
\label{params}

The $uvby\beta$ and Geneva photometry provide convenient means of determining the effective temperature and surface gravity for A and late-B main-sequence stars \citep{moon:1985,kunzli:1997}. Photometric data in both systems are available for \hd. However, it is well-known that standard photometric calibrations, established using normal model atmospheres and non-peculiar stars, may introduce a systematic bias when applied to Ap stars \citep{netopil:2008}. This is due to an enhanced metal line blanketing in their atmospheres, which leads to a redistribution of flux from UV to optical and a significant modification of many photometric parameters. This problem is commonly alleviated by applying empirical corrections to the normal photometric calibrations or deriving calibrations suitable specifically for Ap stars \citep[e.g.][]{stepien:1989,hauck:1993}. However, this approach is obviously inadequate when dealing with such an extreme object as \hd.

Instead, we took advantage of the recent development of realistic model atmospheres of Ap stars capable of taking into account both the non-solar chemical composition of the atmosphere and a strong magnetic field. Specifically, we calculated a grid of model atmospheres and spectral energy distributions covering the \teff\,=\,9000--11000~K and \lgg\,=\,3.5--4.5 ranges with the help of the {\sc llmodels} code \citep{shulyak:2004}. In these calculations we assumed the chemical composition of \hd\ reported by \citet{elkin:2010} and adopted the 27~kG radial magnetic field, corresponding to the phase-averaged mean field modulus of \hd. The Zeeman splitting of spectral lines and polarised radiative transfer were treated in full detail, as discussed by \citet{khan:2006a}. The VALD database \citep{piskunov:1995,kupka:1999} was used as a source of the line opacity data.

Using this set of model atmospheres and the corresponding spectrum energy distributions, we calculated theoretical parameters for the Str\"omgren and Geneva photometric systems and compared them with the observed values. This led to \teff\,=\,10430~K, \lgg\,=\,4.40, and \teff\,=\,10150~K, and \lgg\,=\,4.38 for the $a_0,r^*$ and $c_1,\beta$ calibrations of the $uvby\beta$ system \citep[see][]{moon:1985}. On the other hand, the $pT,pG$ Geneva photometric parameters \citep{kunzli:1997} indicate \teff\,=\,10320~K and \lgg\,=\,4.12. The $B2-G$ Geneva colour suggests \teff\,=\,10030~K for \lgg\,=\,4.3. The hydrogen Balmer lines in our HARPS spectra of \hd\ favour \lgg\,=\,4.3 over \lgg\,=\,4.1. As a compromise, we adopted \teff\,=\,$10250\pm250$~K and \lgg\,=\,$4.3\pm0.1$. 

The broad-band photometric variation of \hd\ has an amplitude of only 10~mmag in Johnson V \citep{freyhammer:2008}. As we show below, the surface gradients of chemical elements (Si, Cr, Fe) that are most important for atmospheric structure are relatively mild. Therefore, it is unlikely that the snapshot Str\"omgren and Geneva photometric measurements used here for \teff\ and \lgg\ determination are significantly affected by stellar variability.

According to the Padova evolutionary tracks \citep{bertelli:2009}, the atmospheric parameters of \hd\ are compatible with a 2.2--2.7\,$M_\odot$ star that has not evolved significantly off the zero-age main sequence. The corresponding stellar radius lies in the range from 1.8 to 2.4\,$R_\odot$. Taking into account the rotational period $P_{\rm rot}=4.048267$~d and the projected rotational velocity \vs\,=\,11.5~\kms\ determined below, the oblique rotator relation yields an inclination angle of the stellar rotational axis $i$\,=\,25--31\degr\ for the aforementioned radius range. In comparison, the $P_{\rm rot}$ uncertainty is completely negligible for $i$ determination, while the \vs\ uncertainty contributes 1.3\degr\ to the total error budget.

\begin{figure}[!th]
\centering
\figps{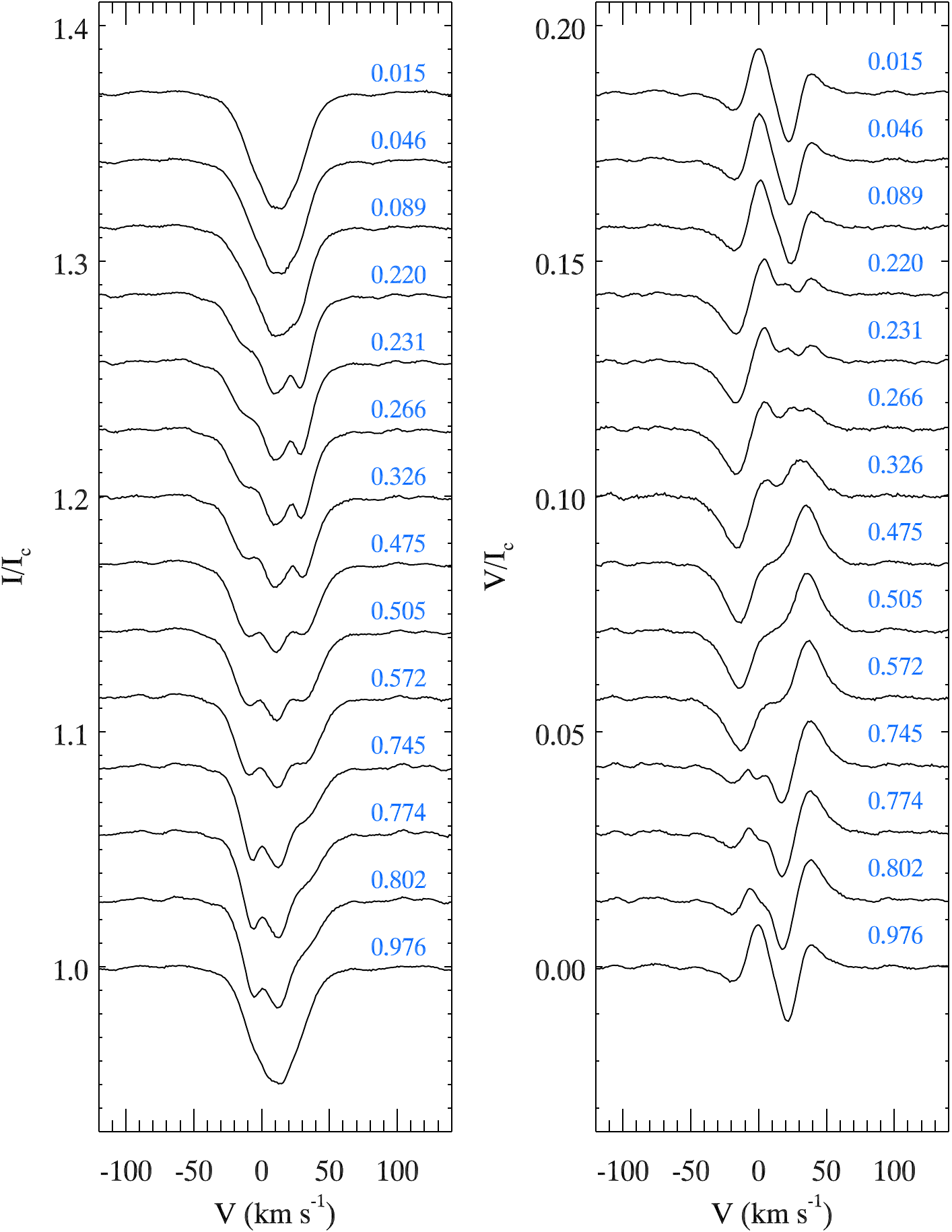}
\caption{Stokes $I$ (left) and $V$ (right) LSD profiles of \hd. The spectra are offset vertically according to the rotational phase indicated in the panels.}
\label{fig:lsd}
\end{figure}

\begin{table}[!th]
\centering
\caption{Mean longitudinal magnetic field of \hd\ measured from LSD profiles. 
\label{tbl:bz}}
\begin{tabular}{ccr}
\hline\hline
HJD & Phase & \bz\ (G)~~ \\
\hline
2455204.8214 &   0.7454 & $-4635\pm39$   \\
2455206.7875 &   0.2311 & $-4317\pm39$   \\
2455207.7753 &   0.4751 & $-9244\pm47$ \\
2455209.8043 &   0.9763 &  $-793\pm40$    \\
2455210.7909 &   0.2200 & $-3987\pm39$     \\
2455316.4739 &   0.3258 & $-6959\pm49$     \\
2455317.4701 &   0.5718 & $-8916\pm48$     \\
2455319.5644 &   0.0892 & $-1245\pm42$    \\
2455601.7825 &   0.8025 & $-3257\pm39$    \\
2455602.7695 &   0.0463 &  $-871\pm39$    \\
2456014.5916 &   0.7743 & $-3878\pm40$   \\
2456015.5662 &   0.0150 &  $-781\pm39$    \\
2456016.5819 &   0.2659 & $-5171\pm41$    \\
2456017.5481 &   0.5046 & $-9051\pm46$    \\
\hline
\end{tabular}
\end{table}

\begin{figure}[!th]
\centering
\figps{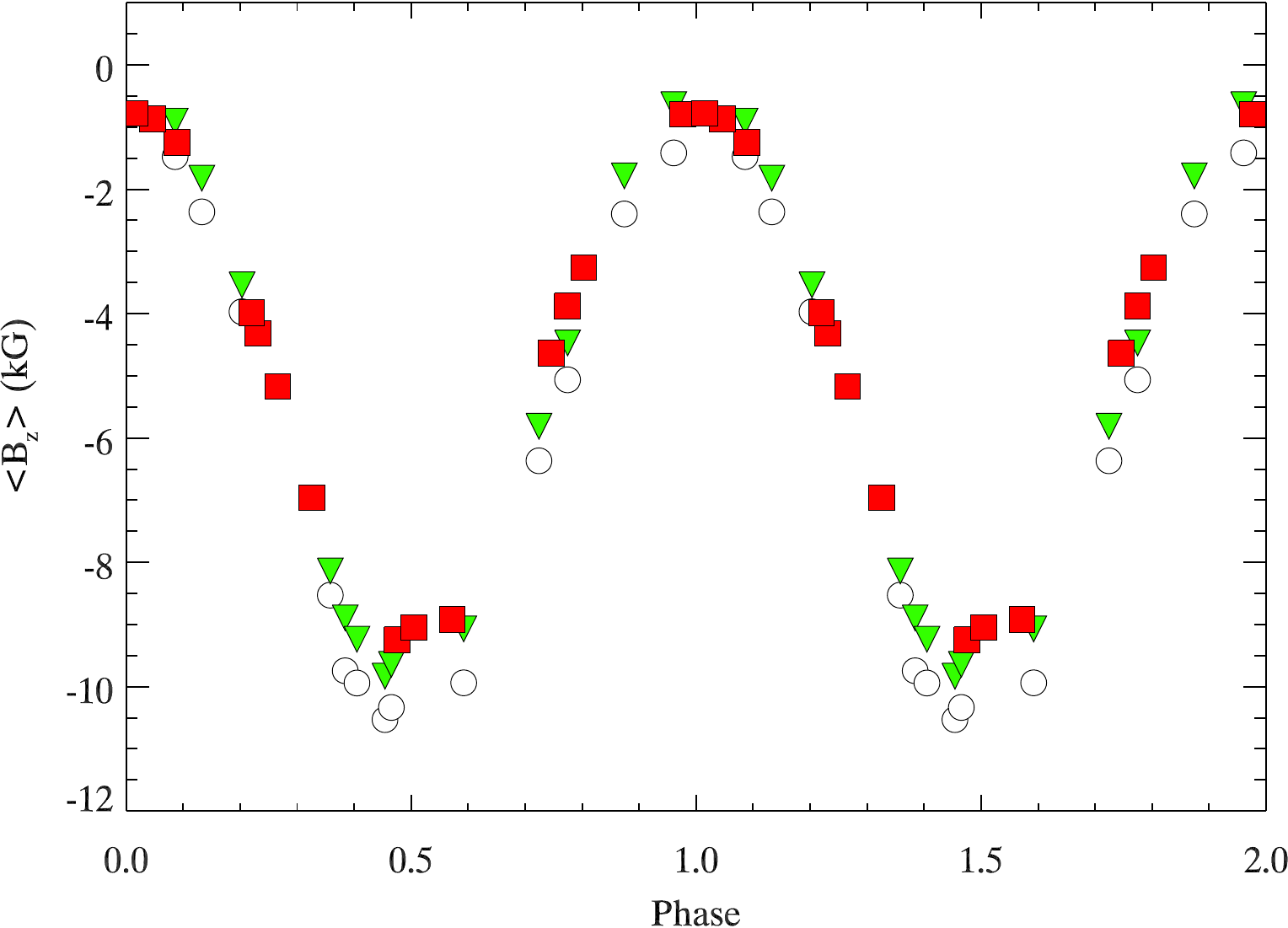}
\caption{Variation of the mean longitudinal magnetic field of \hd. Our measurements from LSD profiles (squares) are compared with \bz\ obtained by \citet{elkin:2010} from metal lines (triangles) and hydrogen lines (circles) using low-resolution FORS1 spectra. In all cases the formal error bars are smaller than the symbol size.}
\label{fig:bz}
\end{figure}

\begin{figure}[!th]
\centering
\figps{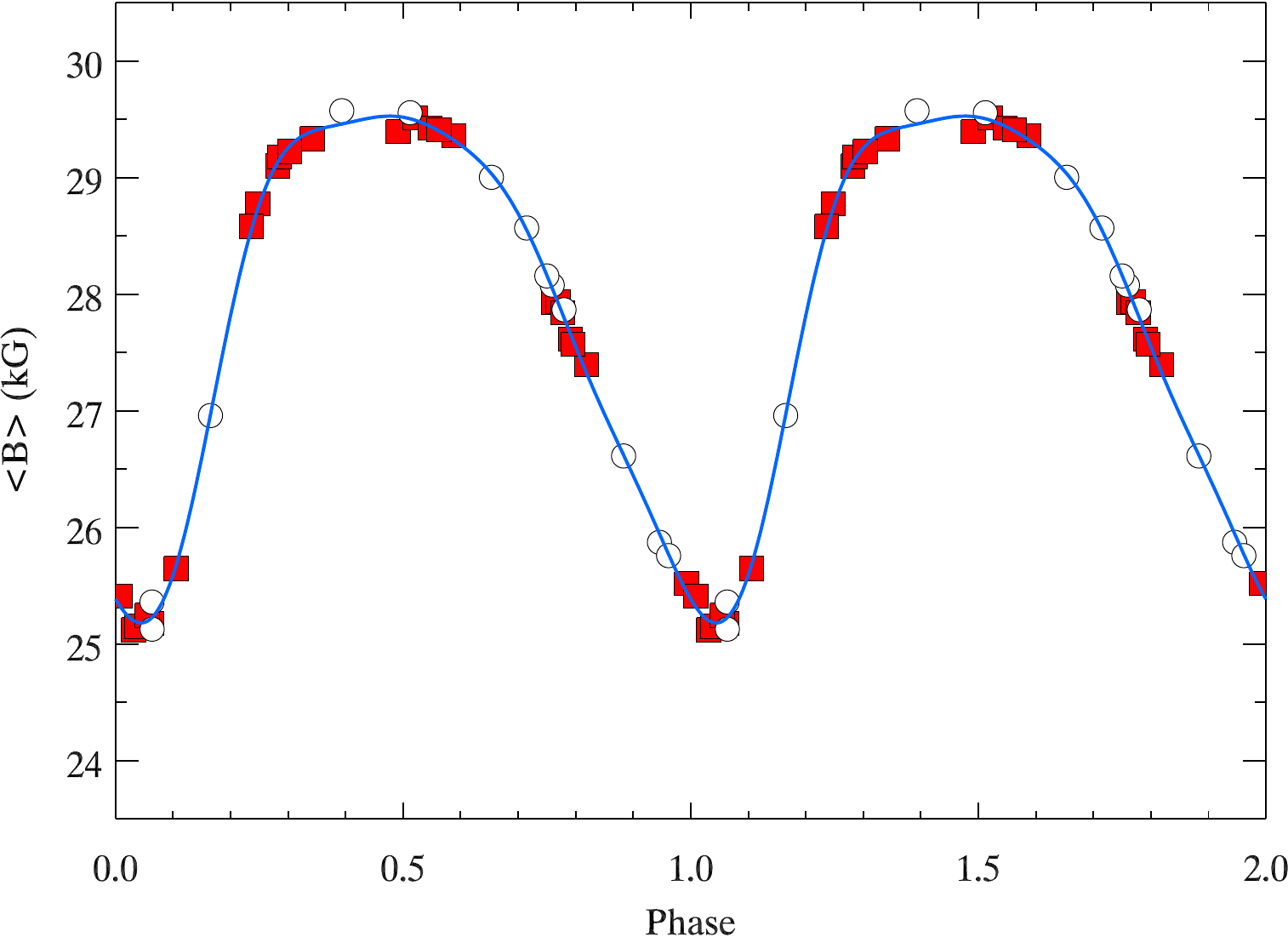}
\caption{Variation of the mean field modulus of \hd\ inferred from the Zeeman resolved components of the \ion{Fe}{ii} 5018.44~\AA\ line. The measurements from HARPS (squares) and UVES (circles) are plotted together with the fourth-order Fourier fit (solid line).}
\label{fig:bm}
\end{figure}

\section{Integral magnetic observables}
\label{moments}

\subsection{Longitudinal magnetic field}

The mean longitudinal magnetic field \bz\ characterises the disk-averaged line-of-sight component of the stellar magnetic field. This is the most commonly used magnetic observable for early-type stars. Longitudinal field can be inferred both from high- and low-resolution circular polarisation spectra \citep{mathys:1991,bagnulo:2002a}. Despite the strong circular polarisation signal in many spectral lines, it is not trivial to determine \bz\ for \hd\ because the line profiles are strongly distorted by the Zeeman splitting and are often severely blended. In addition, many techniques for determining \bz\ rely on the weak-field assumption, which is evidently inappropriate for \hd. Nonetheless, we found that applying the least-squares deconvolution \citep[LSD,][]{donati:1997} to the spectra of our target provides a set of high-precision mean Stokes $I$ and $V$ profiles (see Fig.~\ref{fig:lsd}), which can then be used to establish \bz\ as discussed in \citet{kochukhov:2010a}. These longitudinal field measurements and associated formal error bars are reported in Table~\ref{tbl:bz}. The \bz\ variation of \hd\ is roughly sinusoidal, without a sign reversal and with an extremum at $-9.2$~kG. This is similar to the $-9$~kG longitudinal field found in NGC 2244-334 \citep{bagnulo:2004} and is exceeded only by 12--23~kG \bz\ measured for HD\,215441 \citep{borra:1978}.

The LSD technique indirectly relies on the weak-field assumption, only through a choice of weights used in the line-averaging procedure. With the help of synthetic spectrum calculations we assessed the magnitude of systematic errors that are likely to be present when applying LSD to the polarisation profiles of \hd. Specifically, we computed a grid of synthetic Stokes $V$ spectra for the entire wavelength region covered by the observations and then derived LSD profiles using the same set of lines as in the analysis of the observed spectra. Then we obtained \bz\ from the resulting synthetic LSD profiles and compared these measurements with the input longitudinal field values. This analysis allows us to conclude that the shape of \bz\ curve is unaffected, but the amplitude is underestimated by 5--10\%. 

In the previous study of \hd,\ \citet{elkin:2010} determined the mean longitudinal magnetic field from the low-resolution FORS1 spectra using a correlation between the Stokes $V$ signal and the derivative of Stokes $I$. This approach is different from our methodology and is also formally valid only for weak fields. Nevertheless, as illustrated in Fig.~\ref{fig:bz}, our LSD \bz\ measurements agree very well with the longitudinal field estimated by \citet{elkin:2010} from the metal line regions. On the other hand, their \bz\ curve derived from the hydrogen Balmer lines is offset by about 0.5~kG from their metal line measurements and our LSD results.

\subsection{Mean field modulus}

The mean magnetic field modulus \bm\ is another common Ap-star magnetic observable. It is a measure of the magnetic field strength averaged over the stellar disk. The field modulus can be derived from differential broadening of spectral lines with different magnetic sensitivity \citep[e.g.][]{kochukhov:2006b} or from separation of the Zeeman resolved $\pi$ and $\sigma$ components \citep{mathys:1990}. The latter approach is essentially model independent but requires a combination of a particularly strong magnetic field and slow stellar rotation.

The spectrum of \hd\ exhibits a host of spectral lines with fully resolved Zeeman splitting. However, many of these lines are significantly blended or have complex splitting patterns. The Zeeman components of the \ion{Fe}{ii} 5018.44~\AA\ line appear most suitable for precise \bm\ determination since this line is strong, has a simple triplet-like splitting pattern, and is free from major blends.

The usual method of inferring \bm\ consists of fitting a superposition of Gaussian profiles to individual Zeeman components and applying the formula
\beq
\Delta\lambda_\sigma = 9.337\times10^{-13}\bar{g}\lambda^2_0\langle B \rangle,
\eeq
where wavelength is measured in \AA, \bm\ in Gauss, and $\bar{g}$ is the effective Land\'e factor of the line ($\bar{g}=1.935$ for \ion{Fe}{ii} 5018.44~\AA). The left-hand side of this equation represents the distance between the centres of blue- and red-shifted $\sigma$ components, $\Delta\lambda_\sigma=\lambda_{\sigma_{\rm r}} - \lambda_{\sigma_{\rm b}}$. 

\begin{table}[!th]
\centering
\caption{Mean magnetic field modulus of \hd\ determined from the \ion{Fe}{ii} 5018.44~\AA\ line. 
\label{tbl:bm}}
\begin{tabular}{cccl}
\hline\hline
HJD & Phase & \bm\ (kG) & Instrument \\
\hline
 2454171.5033 & 0.5122 & 29.56 & UVES \\
 2454387.8159 & 0.9456 & 25.87 & UVES \\
 2454432.8249 & 0.0637 & 25.13 & UVES \\
 2454451.8337 & 0.7592 & 28.08 & UVES \\
 2454455.7023 & 0.7148 & 28.57 & UVES \\
 2454468.8459 & 0.9615 & 25.76 & UVES \\
 2454481.8146 & 0.1651 & 26.96 & UVES \\
 2454513.7889 & 0.0633 & 25.36 & UVES \\
 2454516.6890 & 0.7797 & 27.87 & UVES \\
 2454524.6646 & 0.7498 & 28.16 & UVES \\
 2454544.5161 & 0.6535 & 29.00 & UVES \\
 2454547.5113 & 0.3934 & 29.57 & UVES \\
 2454557.5910 & 0.8833 & 26.61 & UVES \\
 2455316.4739 & 0.3420 & 29.33 & HARPS \\
 2455317.4701 & 0.5881 & 29.36 & HARPS \\
 2455319.5644 & 0.1054 & 25.65 & HARPS \\
 2455601.7825 & 0.8187 & 27.40 & HARPS \\
 2455602.7695 & 0.0625 & 25.18 & HARPS \\
 2455204.8214 & 0.7617 & 27.93 & HARPS \\
 2455206.7875 & 0.2473 & 28.78 & HARPS \\
 2455207.7753 & 0.4913 & 29.39 & HARPS \\
 2455209.8043 & 0.9926 & 25.52 & HARPS \\
 2455210.7909 & 0.2363 & 28.58 & HARPS \\
 2456014.5916 & 0.7905 & 27.61 & HARPS \\
 2456015.5662 & 0.0313 & 25.12 & HARPS \\
 2456016.5819 & 0.2822 & 29.10 & HARPS \\
 2456017.5481 & 0.5208 & 29.51 & HARPS \\
 2455609.6785 & 0.7692 & 27.94 & HARPS \\
 2455610.6482 & 0.0087 & 25.41 & HARPS \\
 2456345.5633 & 0.5469 & 29.43 & HARPS \\
 2456345.6263 & 0.5625 & 29.41 & HARPS \\
 2456347.5494 & 0.0375 & 25.15 & HARPS \\
 2456347.6204 & 0.0551 & 25.25 & HARPS \\
 2456348.5530 & 0.2854 & 29.18 & HARPS \\
 2456348.6241 & 0.3030 & 29.22 & HARPS \\
 2456350.5459 & 0.7777 & 27.84 & HARPS \\
 2456350.6169 & 0.7953 & 27.57 & HARPS \\
\hline
\end{tabular}
\end{table}

In general, a simple symmetric analytical function, such as a Gaussian, cannot provide a perfect fit to $\sigma$ components of the disk-integrated profile of a Zeeman triplet. The reasons for this include the distribution of field strengths at the stellar surface and an interplay between rotational Doppler shifts and Zeeman splitting. These two effects lead to a distortion and asymmetries of the $\sigma$ components. This phenomenon is clearly observed for the \ion{Fe}{ii} 5018.44~\AA\ line in the spectrum of \hd\ at many rotational phases. By experimenting with different analytical functions, we found that the fit to the observed profiles of this line can be greatly improved if one uses an asymmetric Gaussian function with individual FWHM parameters for the blue and red wings. Using this method, we were able to derive a set of precise \bm\ measurements for \hd. These field modulus measurements are reported for our 24 HARPS Stokes $I$ spectra in Table~\ref{tbl:bm}. For consistency, we also applied an asymmetric Gaussian fit to the  \ion{Fe}{ii} 5018.44~\AA\ line in 13 UVES observations of \hd\ obtained by \citet{elkin:2010}. The resulting \bm\ measurements are given in Table~\ref{tbl:bm} as well.

\begin{table}[!th]
\centering
\caption{Spectral lines used for magnetic Doppler imaging. 
\label{tbl:lines}}
\begin{tabular}{llrcrr}
\hline\hline
Ion & $\lambda$ (\AA) & $E_{\rm i}$ (ev) & $\bar{g}$ & $\log gf$ & $\log gf_{\rm cor}$ \\
\hline
\ion{Cr}{ii}  & 4242.364 &  3.871 & 1.182 & $-1.170$ & $-0.681$ \\
\ion{Cr}{ii}  & 4252.632 &  3.858 & 1.205 & $-1.810$ & $-1.931$ \\
\ion{Cr}{ii}  & 4284.188 &  3.854 & 0.522 & $-1.670$ & $-1.861$ \\
\ion{Fe}{ii}  & 4491.397 &  2.856 & 0.425 & $-2.700$ & $-2.819$ \\
\ion{Fe}{ii}  & 4508.280 &  2.856 & 0.505 & $-2.250$ & $-2.670$ \\
\ion{Fe}{ii}  & 4520.218 &  2.807 & 1.345 & $-2.600$ & $-1.741$ \\
\ion{Cr}{ii}  & 4592.049 &  4.074 & 1.205 & $-1.221$ & $-1.423$ \\
\ion{Cr}{ii}  & 4634.070 &  4.072 & 0.510 & $-0.990$ & $-1.113$ \\
\ion{Cr}{ii}  & 4812.337 &  3.864 & 1.498 & $-1.960$ & $-1.813$ \\
\ion{Fe}{ii}  & 4923.921 &  2.891 & 1.700 & $-1.320$ & $-1.517$ \\
\ion{Nd}{iii} & 4927.488 &  0.461 & 1.180 & $-0.800$ & $-0.800$ \\
\ion{Fe}{ii}  & 5018.436 &  2.891 & 1.935 & $-1.220$ & $-1.342$ \\
\ion{Nd}{iii} & 5050.695 &  0.296 & 1.170 & $-1.060$ & $-1.575$ \\
\ion{Si}{ii}  & 5978.930 & 10.074 & 1.167 &  $0.040$ &  $0.223$ \\
\ion{Nd}{iii} & 6145.068 &  0.296 & 0.990 & $-1.330$ & $-1.363$ \\
\ion{Nd}{iii} & 6327.265 &  0.141 & 0.935 & $-1.410$ & $-0.988$ \\
\ion{Si}{ii}  & 6347.109 &  8.121 & 1.167 &  $0.170$ &  $0.070$ \\
\ion{Si}{ii}  & 6371.371 &  8.121 & 1.333 & $-0.040$ & $-0.123$ \\
\hline
\end{tabular}
\tablefoot{The columns give the ion designation, central wavelength, excitation potential of the lower atomic level, effective Land\'e factor, oscillator strength from the VALD data base, and corrected oscillator strength.}
\end{table}

The combined \bm\ curve is presented in Fig.~\ref{fig:bm}. We find that the field strength varies from 25.1 to 29.6~kG. The shape of \bm\ variation of \hd\ deviates noticeably from the single- or double-wave sinusoid expected for an oblique dipolar field. In particular, the field strength maximum around phase 0.5 is flattened and the increase to the maximum occurs over a shorter phase interval than the subsequent decrease. We found that a fourth-order Fourier curve is necessary to reproduce the phase variation of the mean field modulus. The scatter of measurements around this curve is only 55~G, giving an estimate of the internal field measurement precision. With the help of this Fourier fit, we determined an improved stellar rotational period of $P_{\rm rot}$\,=\,$4.048267\pm0.000036$~d. The phases given here were computed with this period and the same reference Julian date, 24454509.550, as used by \citet{elkin:2010}.

We estimated an uncertainty of the rotational period from the chi-square fit by adopting the same weight to all measurements and requiring that the reduced chi-square of the fit equals 1. Our rotational period is formally inconsistent with $P_{\rm rot}$\,=\,$4.04899\pm0.00008$~d reported by \citet{elkin:2010}. These authors inferred the period from the mean longitudinal field measurements obtained with the FORS1 instrument. The random and systematic errors of \bz\ inferred from the FORS1 data are notoriously uncertain \citep{bagnulo:2012,bagnulo:2013}. Recently, \citet{landstreet:2014} re-analysed FORS1 observations of \hd, finding a factor of two larger \bz\ errors than \citet{elkin:2010} and obtaining a much more uncertain rotational period, $P_{\rm rot}$\,=\,$4.047\pm0.003$~d. The latter period is formally consistent with our determination.

\section{Magnetic Doppler imaging}
\label{mdi}

\subsection{Methodology}

We carried out a tomographic reconstruction of the magnetic field topology and chemical abundance distributions in \hd\ with the help of the magnetic Doppler imaging code {\sc Invers10} \citep{piskunov:2002a,kochukhov:2002c}. This code performs a simultaneous and self-consistent magnetic and chemical mapping based on polarised radiative transfer calculations with a realistic model atmosphere. Theoretical Stokes spectra are compared with observations in several Stokes parameters and the surface distributions are adjusted iteratively to reproduce the phase variation of the observed intensity and polarisation profiles.

In the version of the {\sc Invers10} code applied here we used a general spherical harmonic expansion of the stellar magnetic field \citep{kochukhov:2014}. The field is parameterised with a set of harmonic coefficients $\alpha_{\ell,m}$, $\beta_{\ell,m}$, $\gamma_{\ell,m}$, describing the radial poloidal, horizontal poloidal, and horizontal toroidal magnetic field components, respectively. Given a relatively low projected rotational velocity of \hd, we limited the harmonic expansion to terms with $\ell \le 5$. In addition, we experimented with inversions assuming a purely dipolar poloidal topology, which in this formalism is realised by setting $\ell=1$, $\alpha=\beta$, and $\gamma=0$. 

A penalty function, designed to minimise unnecessary contribution of higher-order modes \citep[see][]{kochukhov:2014}, was employed in the magnetic inversions with $\ell>1$. At the same time, the standard Tikhonov regularisation \citep{piskunov:2002a} was used for the reconstruction of chemical abundance maps.

For simultaneous magnetic and chemical abundance mapping we selected a set of 14 spectral lines of Si, Cr, and Fe. The main criteria for the line selection was the presence of a clear Stokes $V$ signature at most rotational phases and the absence of significant blending. We used lines with large Land\'e factors ($\bar{g} \ge 1.5$) and lines with low magnetic sensitivity ($\bar{g} \le 0.5$). The latter are useful for constraining \vs\ and improving the reconstruction of the chemical spot distributions. 

Parameters of the spectral lines employed for MDI are listed in Table~\ref{tbl:lines}. This table also includes information on four \ion{Nd}{iii} lines used to reconstruct the abundance distribution of this element from the Stokes $I$ spectra.

A set of oscillator strengths extracted from the VALD data base inevitably contains some errors. Vertical chemical stratification and non-LTE effects, both ignored in our study, may introduce additional discrepancies between relative strengths of spectral lines with different excitation potentials and equivalent widths \citep{ryabchikova:2002,mashonkina:2005}. To remove this line-to-line scatter, we corrected for the oscillator strengths by reducing them to a common system for each ion. In practice, this was accomplished by performing preliminary inversion with a common magnetic field but individual homogeneous abundance distributions for each line. The resulting individual abundances were averaged and deviations from the mean value were adopted as oscillator strength corrections to individual lines. The corrected oscillator strengths are reported in the last column of Table~\ref{tbl:lines}.

The projected rotational velocity was determined by examining the goodness of MDI fits to the lines of Si, Cr, and Fe for different \vs\ values. All three elements consistently point to \vs\,=\,$11.5\pm0.5$~\kms.

According to the discussion in Sect.~\ref{params}, the inclination angle of the stellar rotational axis is probaby at between 25 and 30\degr. We adopted $i=30\degr$ for all inversions discussed below. A variation of $i$ within 5--10\degr\ has a negligible impact on the inversion results.

\begin{figure*}[!th]
\centering
\figgps{16cm}{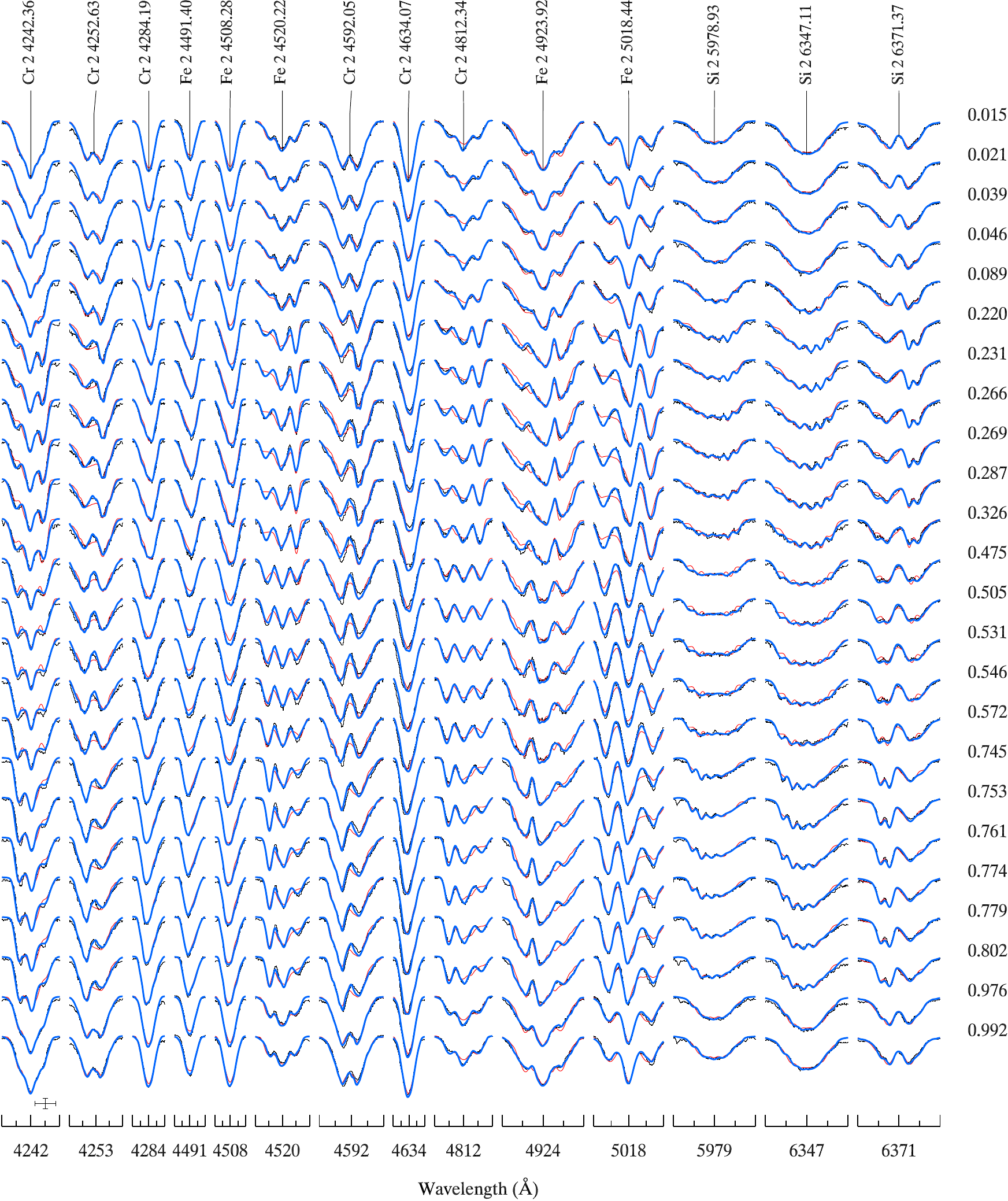}
\caption{Comparison of the observed Stokes $I$ profiles (black line) of Si, Cr, and Fe lines with the fit achieved by the magnetic inversion code using all harmonic modes with $\ell \le 5$ (thick blue line). The thin red line illustrates an attempt to reproduce observations with a dipolar field topology. Spectra corresponding to different rotation phases are offset vertically. Rotation phases are indicated to the right of each spectrum. The error bars in the lower left corner indicate the vertical (10\%) and horizontal (0.5~\AA) scales.}
\label{fig:prfI}
\end{figure*}

\begin{figure*}[!th]
\centering
\figgps{16cm}{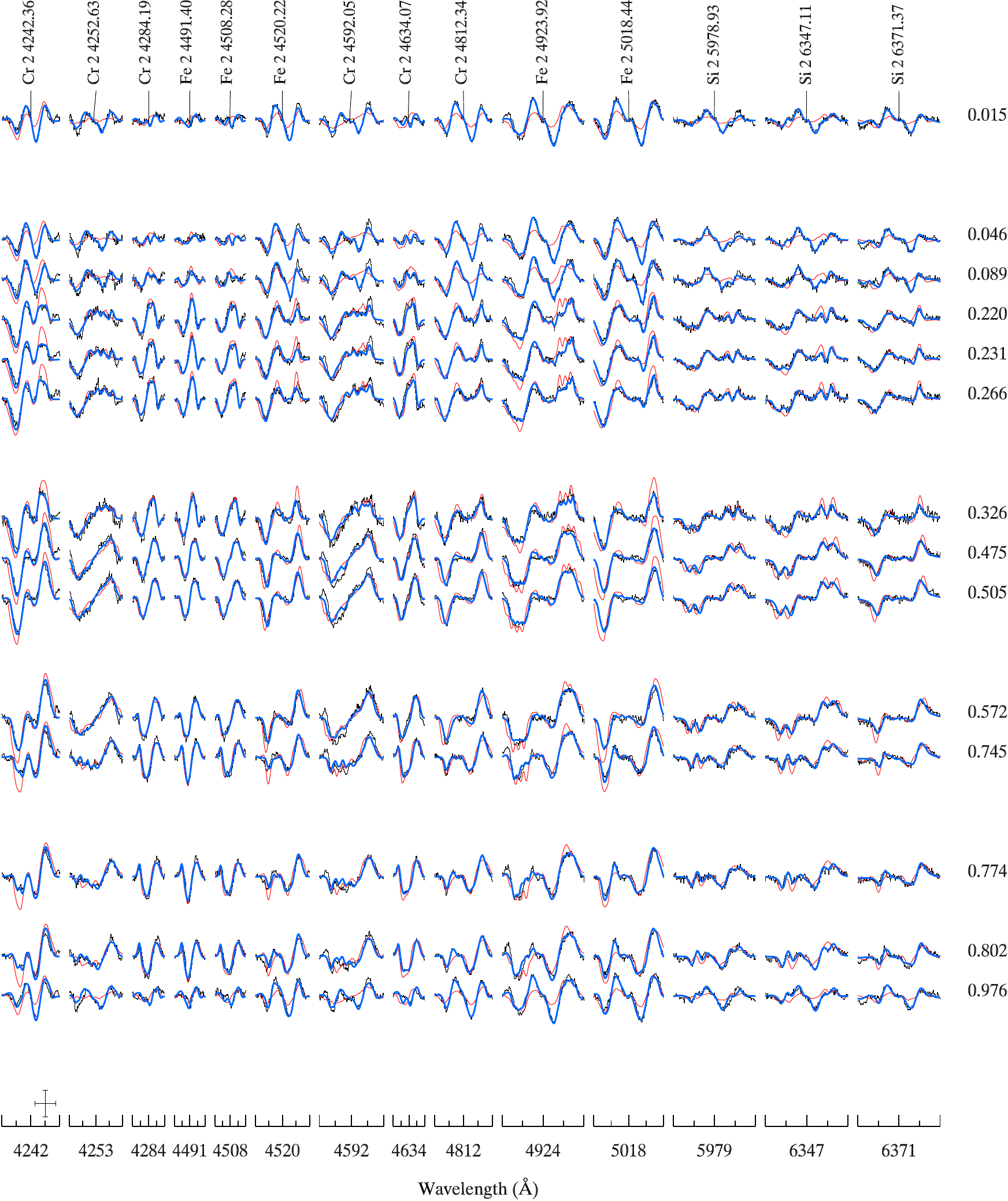}
\caption{Same as Fig.~\ref{fig:prfI} for the Stokes $V$ profiles of Si, Cr, and Fe lines.}
\label{fig:prfV}
\end{figure*}

\begin{figure*}[!th]
\centering
\firrps{14.5cm}{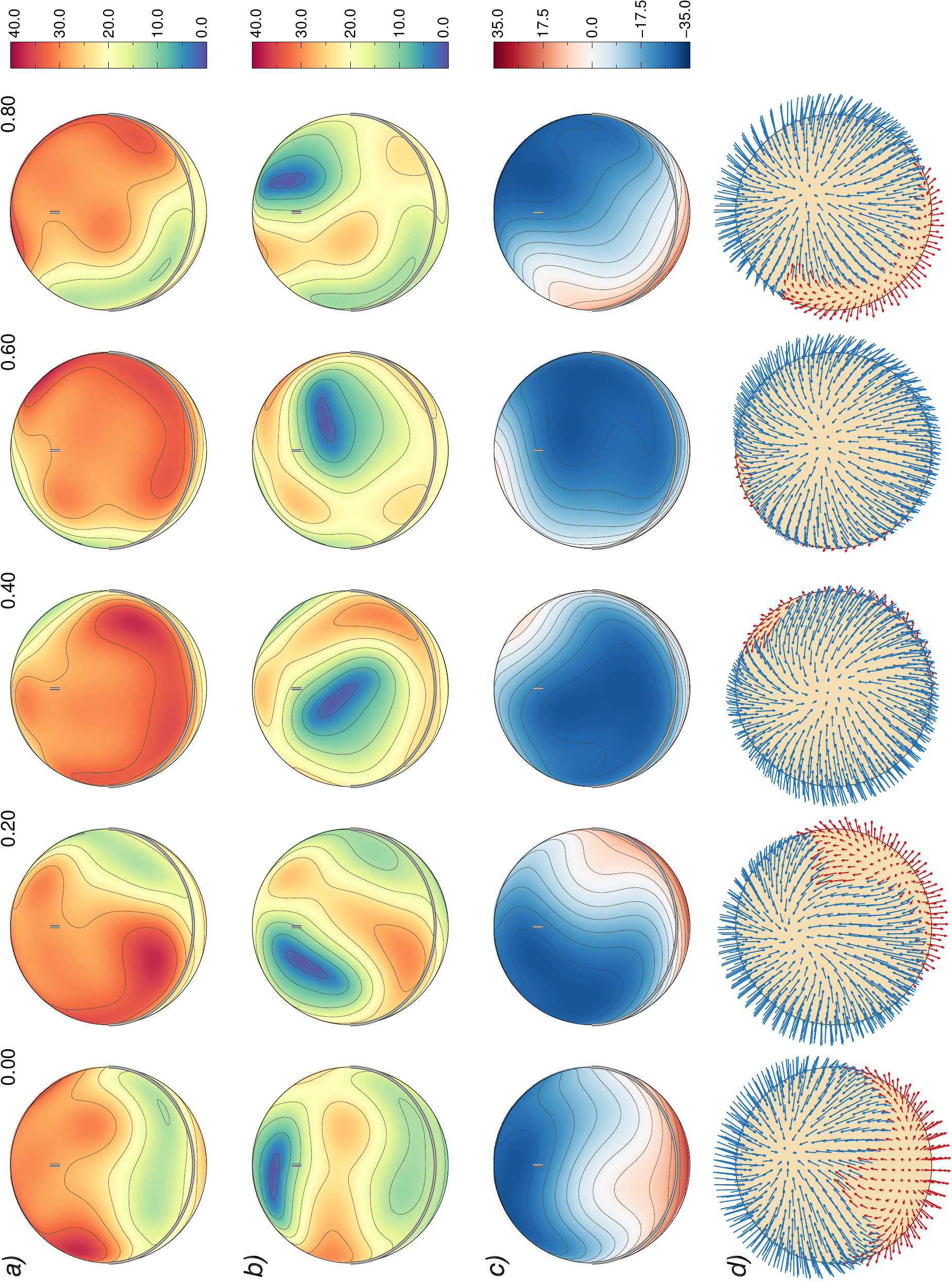}
\caption{Surface magnetic field distribution of \hd\ derived from the Stokes $IV$ profiles of Si, Cr, and Fe lines using MDI. The star is shown at five rotational phases, which are indicated above the spherical plots, and at an inclination angle of $i=30\degr$. The spherical plots show the maps of {\bf a)} the field modulus, {\bf b)} the horizontal field, {\bf c)} the radial field, and {\bf d)} the field orientation. The contours over spherical maps are plotted in steps of 5~kG. The thick line and the vertical bar indicate positions of the rotational equator and the pole. The colour side bars give the field strength in kG. The two different colours in the field orientation map correspond to the field vectors directed outwards (red) and inwards (blue).}
\label{fig:fld}
\end{figure*}

\begin{figure*}[!th]
\centering
\firrps{14.5cm}{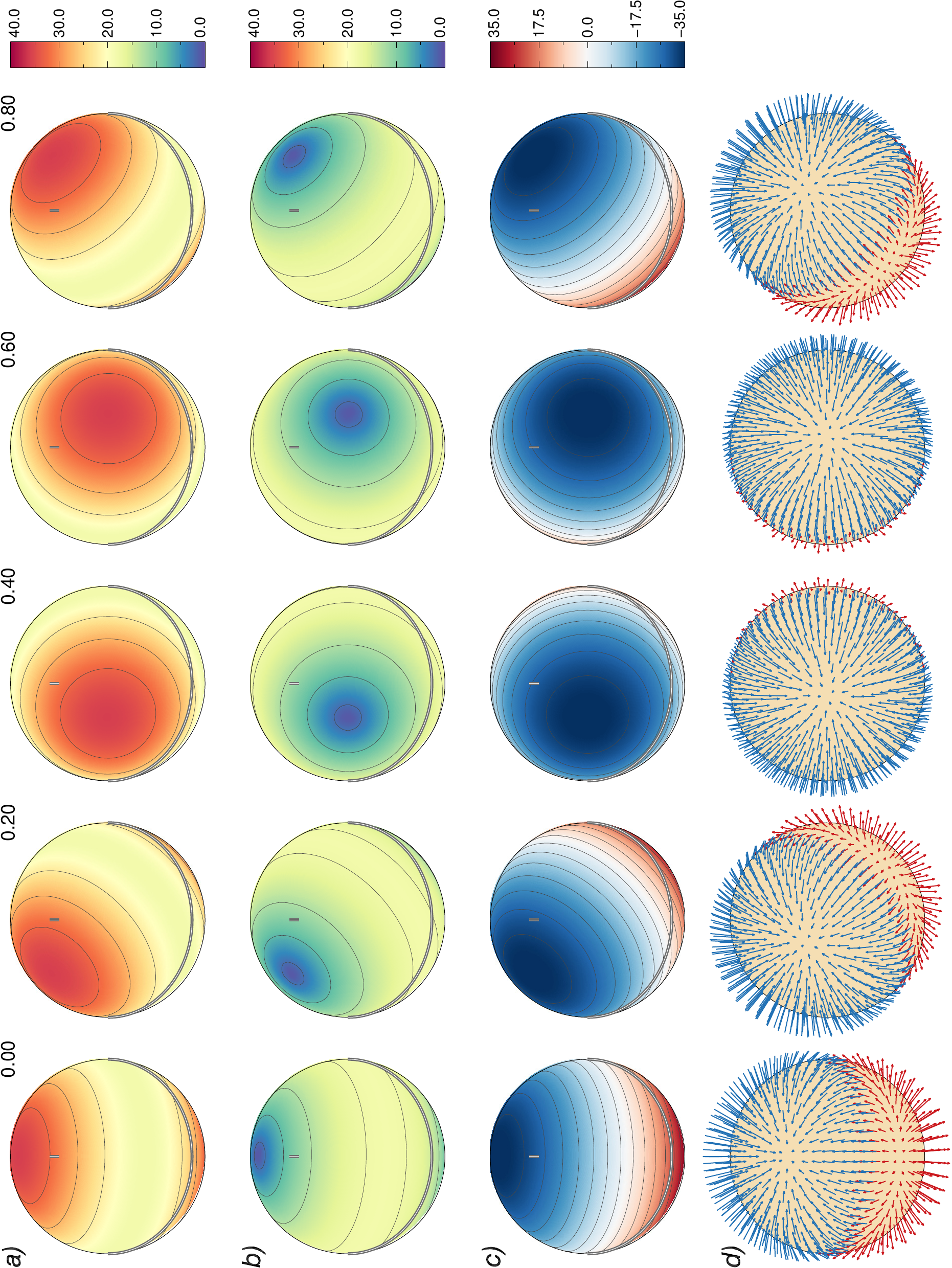}
\caption{Same as Fig.~\ref{fig:fld} for the best-fit dipole field.}
\label{fig:fld_dipole}
\end{figure*}

\begin{figure*}[!th]
\centering
\firrps{14.5cm}{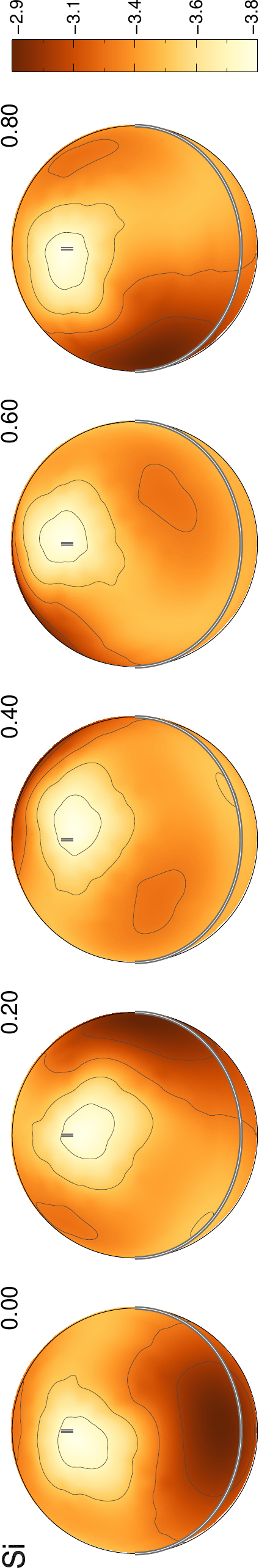}\\\vspace*{0.5cm}
\firrps{14.5cm}{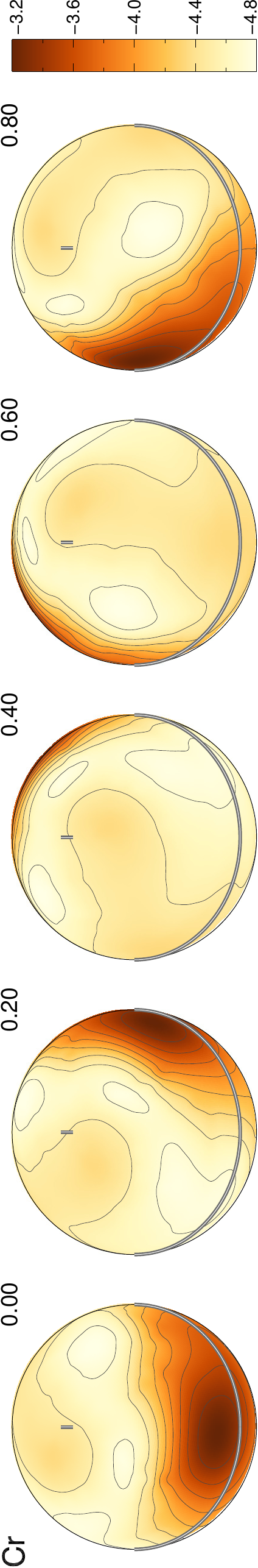}\\\vspace*{0.5cm}
\firrps{14.5cm}{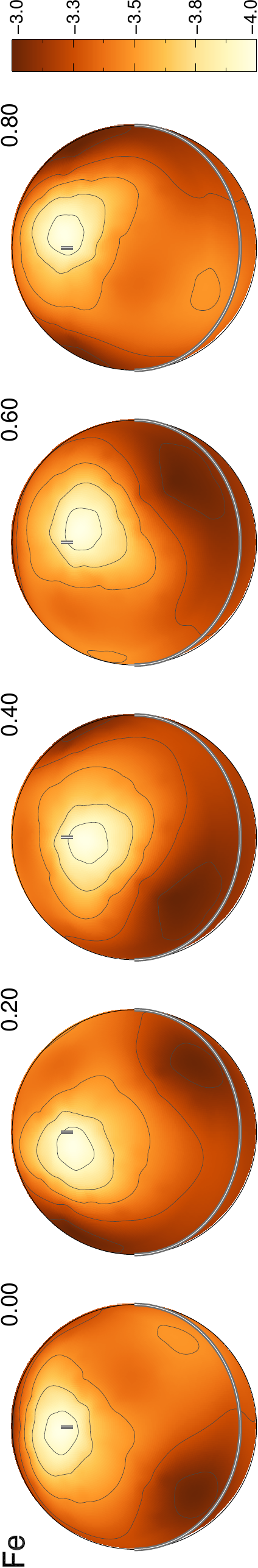}\\\vspace*{0.5cm}
\firrps{14.5cm}{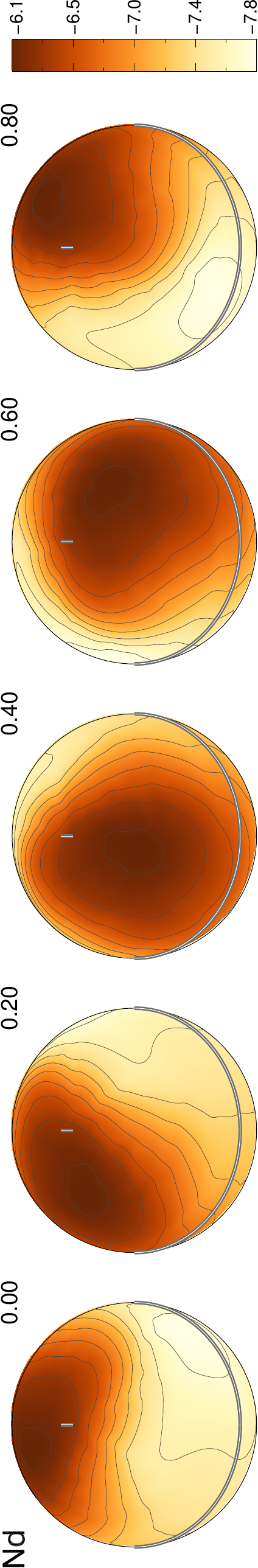}
\caption{Chemical spot distributions of Si, Cr, Fe, and Nd reconstructed with MDI. The star is shown at five rotational phases, as indicated next to each plot. The contours over spherical maps are plotted with a 0.2~dex step. The side bars give element abundances in logarithmic units $\log (N_{\rm el}/N_{\rm tot})$.}
\label{fig:abn}
\end{figure*}

\subsection{Magnetic field topology}

\begin{figure}[!th]
\centering
\figgps{8cm}{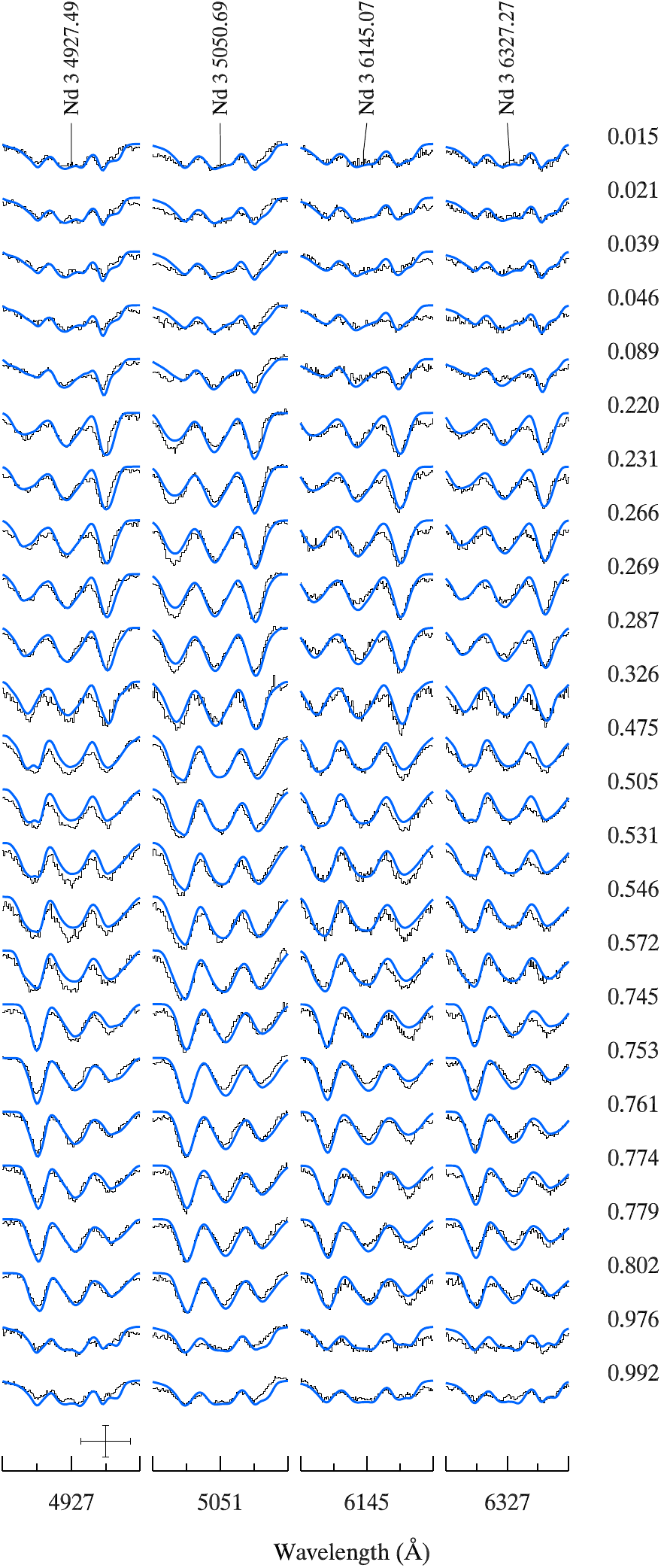}
\caption{Comparison of the observed Stokes $I$ profiles (black line) of Nd lines with the fit achieved by the magnetic inversion code (thick blue line). The format of this figure is similar to that of Fig.~\ref{fig:prfI}.}
\label{fig:prfNd}
\end{figure}

The field topology of \hd\ was derived with the help of \mbox{{\sc Invers10}} by fitting the phase variation of the Stokes $I$ and $V$ spectra of the 14 Si, Cr, and Fe lines listed in Table~\ref{tbl:lines}. Chemical abundance distributions were derived simultaneously for all three elements. Results of the inversions are presented in Figs.~\ref{fig:prfI}--\ref{fig:abn}. The first pair of figures compares HARPSpol observations with the theoretical intensity (Fig.~\ref{fig:prfI}) and circular polarisation (Fig.~\ref{fig:prfV}) profiles. The next two figures illustrate the inferred surface distribution of magnetic field for the inversion with a general harmonic parameterisation (Fig.~\ref{fig:fld}) and for the restricted dipolar fit (Fig.~\ref{fig:fld_dipole}). The different rows in these figures show spherical projections of the surface distributions for the field modulus, the total horizontal field, and the radial field. The bottom row shows a vector field map that includes highlighted regions of opposite polarity. Finally, Fig.~\ref{fig:abn} shows the spherical maps of the chemical abundance distributions obtained in our study.

As evident from Figs.~\ref{fig:prfI} and \ref{fig:prfV}, the inversion code successfully reproduces all available observations, both in Stokes $I$ and $V$. The resulting magnetic map is best described as a distorted dipolar field. In agreement with the longitudinal field variation, large areas of both negative (inward directed field) and positive (outward directed field) polarities are present on the stellar surface at the magnetic minimum phase ($\varphi\approx0$). On the other hand, at the magnetic maximum around $\varphi\approx0.5$ the entire visible stellar surface is covered by a negative field. 

The phase variation of the relative intensities of the $\pi$ and $\sigma$ components in the Zeeman-resolved spectral lines is also broadly compatible with a dipolar behaviour. The strength of the $\sigma$ components is proportional to $1+\cos^2\theta$, where $\theta$ is the angle between the local field vector and the line of sight. At phase 0 we observe the magnetic equator with large amounts of transverse field. In accordance with this, Zeeman triplet lines exhibit weaker $\sigma$ components. Conversely, at phase 0.5, when the field is predominantly directed along the line of sight, the $\sigma$ components become stronger.

According to the maps presented in Fig.~\ref{fig:fld}, the maximum field modulus reached at the stellar surface is 39~kG and the strongest local radial field is $-29$~kG. In terms of the energy contributions of different harmonic terms, the magnetic field of \hd\ is almost entirely poloidal (96\% of the total magnetic energy) and dipolar (90\% of the magnetic energy is concentrated in $\ell=1$ modes). 

But despite the general dipolar-like behaviour, our inversion results indicate that the field of \hd\ deviates from a pure oblique dipole in a number of subtle aspects. For example, the region with the strongest local field modulus forms a longitudinally extended arc (see Fig.~\ref{fig:fld}a) rather than a circular spot at the negative pole. There are also numerous deviations from axisymmetry at the level of up to 40\% for the field modulus and 30\% for the radial field maps. An examination of the inversion results reveals that these deviations from the canonical dipolar topology are mainly accomplished not by introducing higher-order multipolar terms, but by modifying the horizontal poloidal field ($\beta$ coefficients in the harmonic expansion) relative to the radial field ($\alpha$ coefficients) for the same $\ell$ and $m$ values.

We have assessed significance of the deviation of the magnetic field geometry of \hd\ from the pure dipolar shape by comparing in detail magnetic inversions for the general multipolar expansion and for the restricted dipolar fit. The optimal dipolar field, illustrated in Fig.~\ref{fig:fld_dipole}, has a polar strength of $B_{\rm p}=36$~kG and an obliquity of $\beta=36\degr$. The corresponding theoretical Stokes profiles are displayed in Figs.~\ref{fig:prfI} and \ref{fig:prfV} with thin red lines. As evident from Fig.~\ref{fig:prfI}, the dipolar model is reasonably successful in reproducing variation of the Stokes $I$ profiles. The only apparent systematic problem is the stronger asymmetry of the $\sigma$ components for the lines with triplet splitting (e.g. \ion{Cr}{ii} 4812.34, \ion{Fe}{ii} 4520.22, 5018.44~\AA). Nevertheless, in terms of the formal fit quality, characterised by the mean deviation between the observed and computed profiles, the dipolar fit provides an about 40\% poorer description of the observed intensity spectra than the inversion with all $\ell\le5$ harmonic terms.

The dipolar model is far less successful in reproducing details of the Stokes $V$ profile variation. Figure~\ref{fig:prfV} shows
that this model only matches the circular polarisation spectra of lines with small Land\'e factors. For all lines with strong Zeeman splitting this model yields stronger $V$ profiles than in observations at phase 0.5. It also predicts some sharp profile details, for instance for \ion{Fe}{ii} 4923.92~\AA, that are not observed. The amplitude of the Stokes $V$ signatures is also significantly underestimated at the magnetic minimum around phase 0.0. The overall quality of the dipolar Stokes $V$ fit is reduced by as much as 65\% compared with the magnetic inversion with a full set of harmonic terms. In conclusion, although departures from a dipolar field geometry of \hd\ are relatively small, they appear to be statistically significant and are justified by the observational data at hand.

\subsection{Chemical abundance distributions}

We reconstructed surface abundance distributions of four chemical elements. The maps of Si, Cr, and Fe were obtained simultaneously with the magnetic field topology, as explained above. In addition, the distribution of Nd was derived from the Stokes $I$ profiles of four \ion{Nd}{iii} spectral lines using the previously inferred magnetic field geometry. Chemical abundance maps are presented in Fig.~\ref{fig:abn}. Figure~\ref{fig:prfNd} compares the observed and computed profiles of \ion{Nd}{iii} lines.

In general, \hd\ does not exhibit large surface chemical gradients. The contrast of both Si and Fe abundance distributions, determined by considering 90\% of surface zones, is about 0.5~dex. Both elements are overabundant by $\sim$\,1~dex on average relative to the Sun and show a relative enhancement close to the rotational phase 0, when the magnetic equatorial region is visible at the stellar surface. The Fe spot at phase 0 is slightly offset relative to similar structure in the Si map. In addition, Fe distribution exhibits a secondary enhancement area around phase 0.5.

Both Si and Fe show a relative underabundance zone at the rotational pole. We suspect this feature to be an artefact of ignoring vertical chemical stratification. Indeed, the concentration of light and iron peak elements is observed to decrease with height in Ap stars \citep{ryabchikova:2002,kochukhov:2006b}. This leads to line profile shapes with extended wings and a shallow core. These anomalies would be interpreted as a pole-to-equator abundance increase by a DI code that assumes a homogeneous vertical distribution of chemical elements.

The surface distributions of Cr and Nd are characterised by higher contrasts: 1.2 and 1.4~dex. The average concentration of these elements in the atmosphere of \hd\ exceeds the corresponding solar abundances by 2.4~dex for Cr and 3.6~dex for Nd. The chromium surface map is dominated by a compact spot situated at the magnetic equator region. Neodymium shows an opposite distribution with a large overabundance area roughly coinciding with the stellar surface region covered by negative magnetic field. Neither element shows the axisymmetric polar feature found in the Si and Fe maps.

\section{Conclusions and discussion}
\label{disc}

We have carried out a detailed analysis of the phase-resolved, very high-resolution intensity and circular polarisation HARPSpol spectra of \hd. This object is, in many respects, an extreme Ap star, with the second-strongest magnetic field known among non-degenerate stars. We have presented new measurements of the longitudinal magnetic field and the mean field modulus for \hd. The HARPS field modulus data, combined with measurements derived by us from archival spectra, were used to refine the stellar rotational period. We also derived atmospheric parameters of \hd\ with the help of modern theoretical model atmospheres, which included a direct treatment of non-solar chemical abundances and the Zeeman effect due to a strong magnetic field. Finally, we performed tomographic mapping of the surface of \hd\ using magnetic Doppler imaging. This analysis yielded a detailed map of the stellar magnetic field topology and distributions of four chemical elements. Both the magnetic field topology analysis and the chemical mapping results presented in our study of \hd\ probe previously unexplored parameter space and thus provide valuable constraints on the magnetohydrodynamic models of the envelopes of Ap/Bp stars. 

Based on the results of our MDI analysis, we conclude that the morphology of the surface features in \hd\ is not qualitatively different from surface inhomogeneities found in Ap stars with much weaker magnetic fields. The field geometry of \hd\ is poloidal and close to dipolar. Deviations from an axisymmetric oblique dipole configuration are small but statistically significant. In this respect, the geometrical properties of the magnetic field in \hd\ are similar to other Ap stars studied with the Stokes $IV$ imaging \citep[e.g.][]{luftinger:2010a,kochukhov:2014},
even though our target has a field that is an order of magnitude stronger. At the same time, the contrast of the chemical abundance distributions appears to be considerably lower in \hd\ than in stars with weaker fields \citep{lueftinger:2003,kochukhov:2004e,freyhammer:2009}.

The analysis of HD\,215441=Babcock's star by \citet{landstreet:1989} is the only previous line profile study of an Ap star with $\sim$\,30~kG mean field strength. In that paper an axisymmetric model of the magnetic field topology and schematic axisymmetric abundance maps were derived by fitting Zeeman-resolved lines in the intensity spectra. The spectral resolution of these data was more than twice as low as in our study, and circular polarisation spectra were not available at all. The field geometry of HD\,215441 was modelled with a combination of axisymmetric dipole, quadrupole, and octupole components of comparable strength. The best-fit magnetic field model found by \citet{landstreet:1989} is roughly bipolar, with considerable asymmetry in the field strength between the visible and invisible magnetic poles. At the same time, the Stokes $I$ line profile fits presented by \citet{landstreet:1989} are generally inferior to those achieved in our analysis of \hd. Together with the lack of high-resolution polarisation profile modelling, this makes details of the magnetic field topology of HD\,215441 somewhat uncertain and leaves some room for improvement. It would be worthwhile to re-analyse this star using modern high-quality observations in several Stokes parameters.

Two main conclusions regarding the field strength dependence of the field topology and the contrast of chemical abundance distributions can be drawn by comparing previous studies and our MDI results obtained for \hd. First, mechanisms that create and sustain global magnetic fields in intermediate-mass main-sequence stars operate independently of the field intensity over at least two orders of magnitude (from $\sim$\,0.3 to $\sim$\,30~kG) of the field strength. Instead, there appears to be a dependence on the stellar mass, with very complex non-dipolar field geometries found exclusively in higher-mass early B-type stars \citep{donati:2006b,kochukhov:2011a}.

Second, if the process of radiative diffusion altered by magnetic field is assumed to be responsible for the chemical inhomogeneities on the surfaces of Ap stars, its efficiency is independent of the field strength in the range between 0.3 and $\sim$\,10~kG, as can be judged from comparable spot contrasts in $\varepsilon$~UMa,
for example \citep[][$B_{\rm p}=0.4$~kG]{lueftinger:2003,bohlender:1990a}, HD\,83368 \citep[][$B_{\rm p}=2.5$~kG]{kochukhov:2004e}, and HD\,3980 \citep[][$B_{\rm p}=6.9$~kG]{nesvacil:2012}. At the same time, according to the results obtained here for \hd\ and by \citet{landstreet:1989} for HD\,215441, the surface abundance contrasts diminish as the field strength reaches several tens of kG.

\begin{acknowledgements}
OK is a Royal Swedish Academy of Sciences Research Fellow, supported by the grants from the Knut and Alice Wallenberg Foundation, the Swedish Research Council, and the G\"oran Gustafsson Foundation. 
The computations presented in this paper were performed on resources provided by SNIC through the Uppsala Multidisciplinary Center for Advanced Computational Science (UPPMAX).
\end{acknowledgements}


\end{document}